\documentclass{kapproc} 

\let\footnote\savefootnote

\usepackage{procps} 

\usepackage[dvips]{graphicx}

\upperandlowercase

\kluwerbib 

\begin{document}

\articletitle{The 2~Ms Chandra Deep Field-North}

\articlesubtitle{Moderate-luminosity AGNs and dusty starburst galaxies}

\author{D.~M. Alexander,\altaffilmark{1} F.~E. Bauer,\altaffilmark{2}
  W.~N. Brandt,\altaffilmark{2} and
  A.~E. Hornschemeier\altaffilmark{3}}

\altaffiltext{1}{Institute of Astronomy, Madingley Road, Cambridge,
  CB3 0HA, UK}

\altaffiltext{2}{Department of Astronomy \& Astrophysics, The
  Pennsylvania State University, PA 16802, USA}

\altaffiltext{3}{Department of Physics and Astronomy, Johns Hopkins
  University, MD 21218, USA}

\begin{abstract}
  The 2~Ms Chandra Deep Field-North survey provides the deepest view
  of the Universe in the 0.5--8.0 keV X-ray band. In this brief review
  we investigate the diversity of X-ray selected sources and focus
  on the constraints placed on AGNs (including binary AGNs) in
  high-redshift submm galaxies.

\end{abstract}


%
\section{Introduction}
%

The 2~Ms Chandra Deep Field-North (CDF-N) survey provides the deepest
view of the Universe in the 0.5--8.0 keV band. It is $\approx$~2 times
deeper than the 1~Ms {\it Chandra} surveys (Brandt et~al. 2001;
Giacconi et~al. 2002) and $\approx$~2 orders of magnitude more
sensitive than pre-{\it Chandra} surveys. Five hundred and three (503)
highly significant sources are detected over the 448~arcmin$^2$ area
of the CDF-N, including 20 sources in the central 5.3~arcmin$^2$
Hubble Deep Field-North region (Alexander et~al. 2003a; see Fig~1).
The on-axis flux limits of
$\approx2.3\times10^{-17}$~erg~cm$^{-2}$~s$^{-1}$ (0.5--2.0~keV) and
$\approx1.4\times10^{-16}$~erg~cm$^{-2}$~s$^{-1}$ \hbox{(2--8~keV)}
are sensitive enough to detect moderate-luminosity starburst galaxies
out to $z\approx$~1 and moderate-luminosity AGNs out to $z\approx$~10.


In addition to deep X-ray observations, the CDF-N region also has deep
multi-wavelength imaging (radio, submm, infrared, and optical) and
deep optical spectroscopy (e.g.,\ Barger et~al. 2003a). Most recently,
the CDF-N has been observed with the ACS camera on {\it HST} and will
be observed with the IRAC and MIPS cameras on {\it SIRTF} as part of
the GOODS project (Dickinson \& Giavalisco 2003). The {\it HST} data,
in particular, are providing key morphological and environmental
constraints on the X-ray detected sources.


%
\section{The diversity of X-ray selected sources}
%

\begin{figure}[t]
\sidebyside
{\centerline{\includegraphics[width=2in]{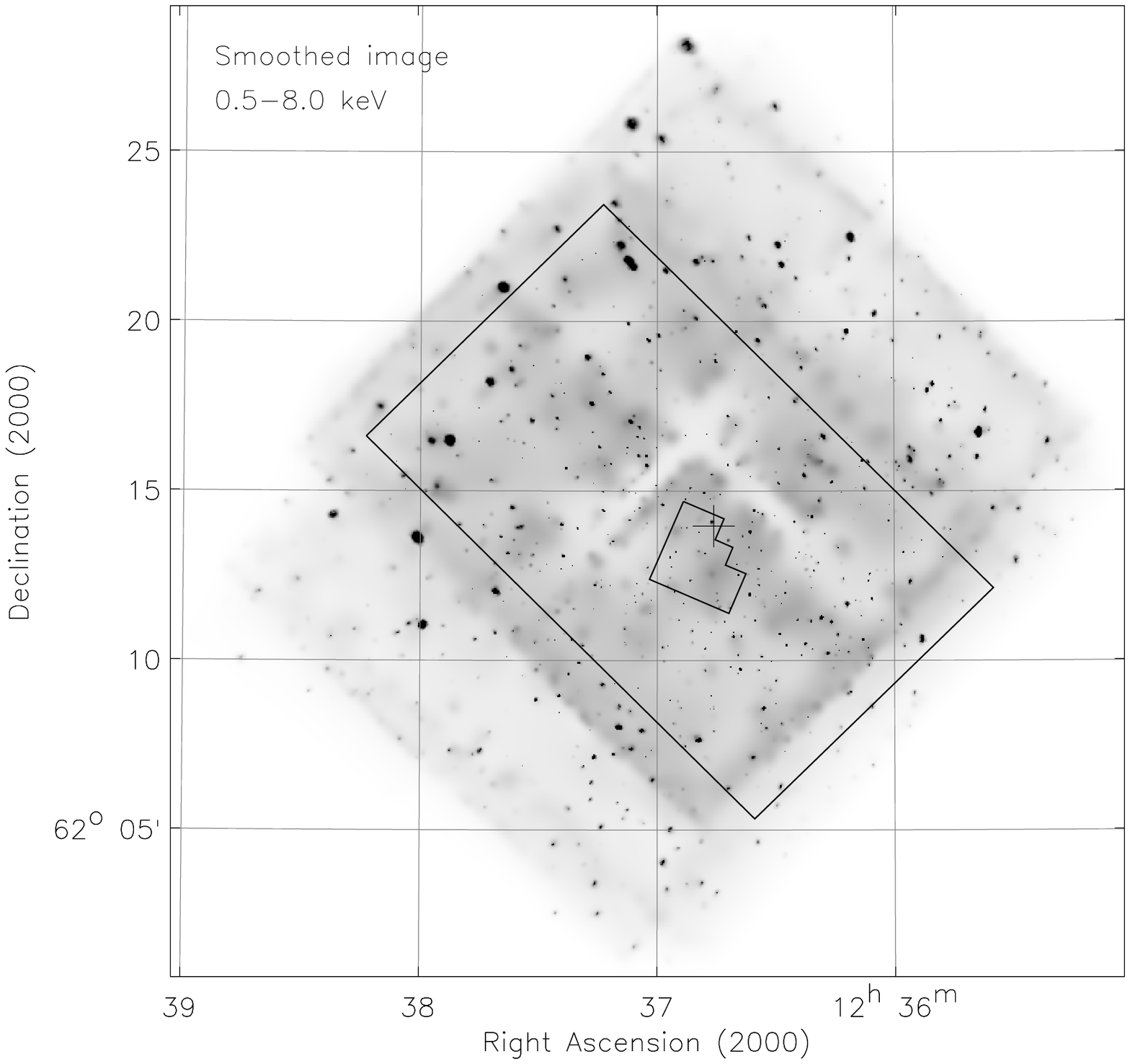}}
\caption{Adaptively smoothed ($2.5\sigma$) 0.5--8.0~keV image of the CDF-N 
  (see Fig~3 of Alexander et~al. 2003a).}}
{\centerline{\includegraphics[width=2in]{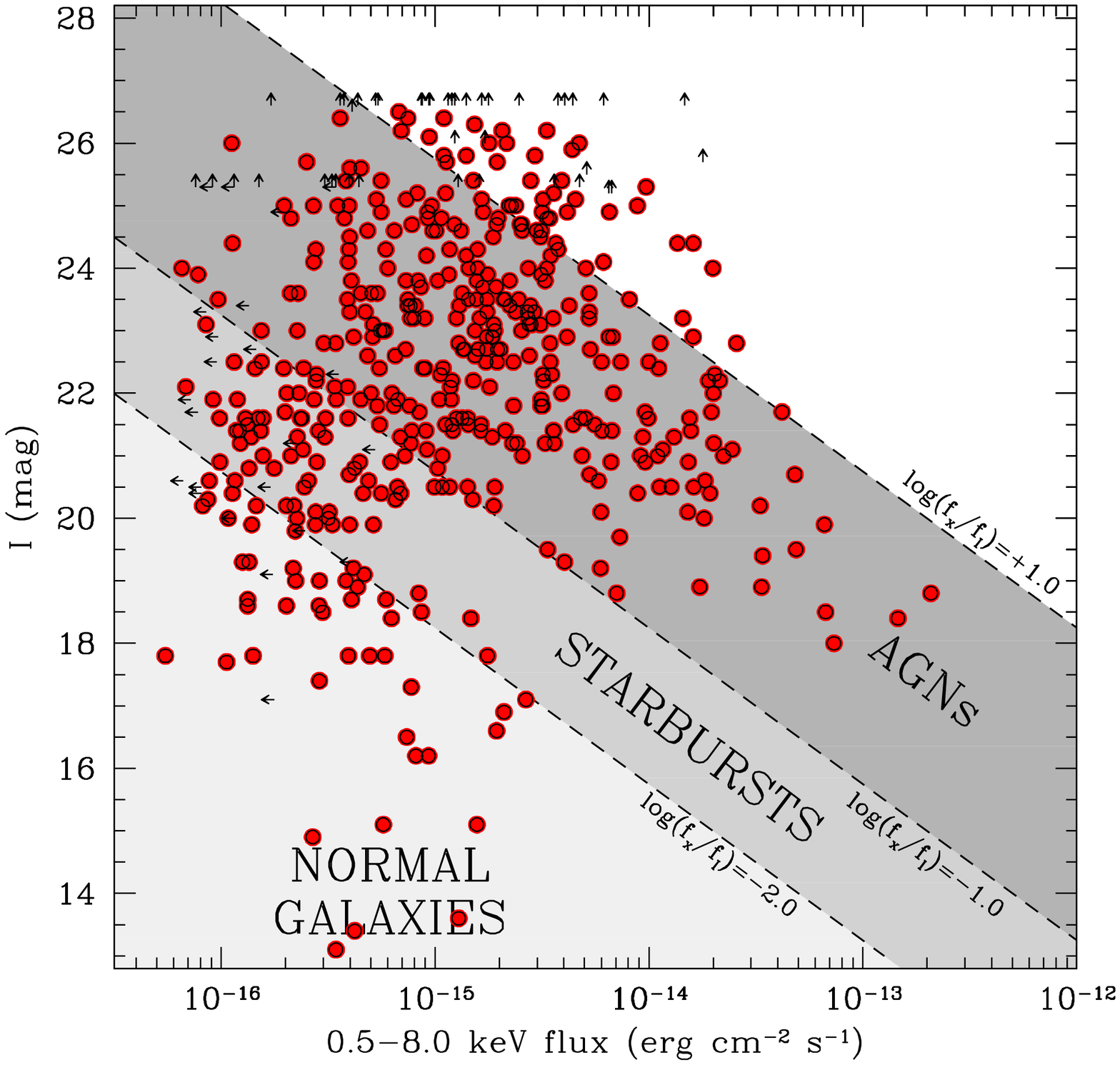}}
\caption{$I$-band magnitude versus X-ray flux. The shaded regions show 
approximate flux ratios for different source types.}}
\end{figure}

The X-ray-to-optical flux ratios of the faintest X-ray sources span up
to five orders of magnitude, indicating a broad variety of source
types (including AGNs and starburst galaxies; see Fig~2). Many of the
AGNs show the optical characteristics of AGN activity (e.g.,\ broad or
highly ionised emission lines). However, a large fraction (perhaps
$>50$\%) are either too faint for optical spectroscopic identification
or do not show typical AGN optical features (e.g.,\ Alexander et~al.
2001; Hornschemeier et~al. 2001; Comastri et~al.  2003). It is the
X-ray properties of these sources that identify them as AGNs [e.g.,\ 
luminous, and often variable and/or hard (i.e.,\ $\Gamma<1$) X-ray
emission].  The corresponding AGN source density ($\approx$~6000
deg$^{-2}$) is $\approx$~10 times higher than that found by the
deepest optical surveys (e.g.,\ Wolf et~al. 2003).

X-ray spectral analyses of the X-ray brightest AGNs indicate that both
obscured and unobscured sources are found (Vignali et~al. 2002; Bauer
et~al. 2003). However, few Compton-thick sources have been identified,
and current analyses suggest that they are rare even at faint X-ray
fluxes (e.g.,\ Alexander et~al.  2003a). Whilst AGNs are identified
out to $z=5.189$ fewer high-redshift moderate-luminosity AGNs are
found than many models predict (e.g.,\ Barger et~al. 2003b). Indeed,
current analyses suggest that moderate-luminosity AGN activity peaked
at comparitively low redshifts (e.g.,\ Cowie et~al. 2003).

A large number of apparently normal galaxies are detected at faint
X-ray fluxes (e.g.,\ Hornschemeier et~al. 2001). The properties of
these sources at infrared, radio, X-ray, and optical wavelengths are
consistent with those expected from starburst and normal galaxies
(e.g.,\ Alexander et~al. 2002; Bauer et~al. 2002; Hornschemeier et~al.
2003). Furthermore, their X-ray and radio luminosities are correlated
in the same manner as for local starburst galaxies, suggesting that
the X-ray emission can be used directly as a star-formation indicator
(Bauer et~al. 2002; Ranalli et~al. 2003; see Fig~3). While the X-ray
emission from the low-redshift, low-luminosity sources could be
produced by a single ultra-luminous X-ray source (e.g.,\ Hornschemeier
et~al. 2003), the majority of these sources have X-ray luminosities
between those of M~82 and NGC~3256, implying moderate-to-luminous
star-formation activity.

\begin{figure}[t]
\sidebyside
{\centerline{\includegraphics[width=2in]{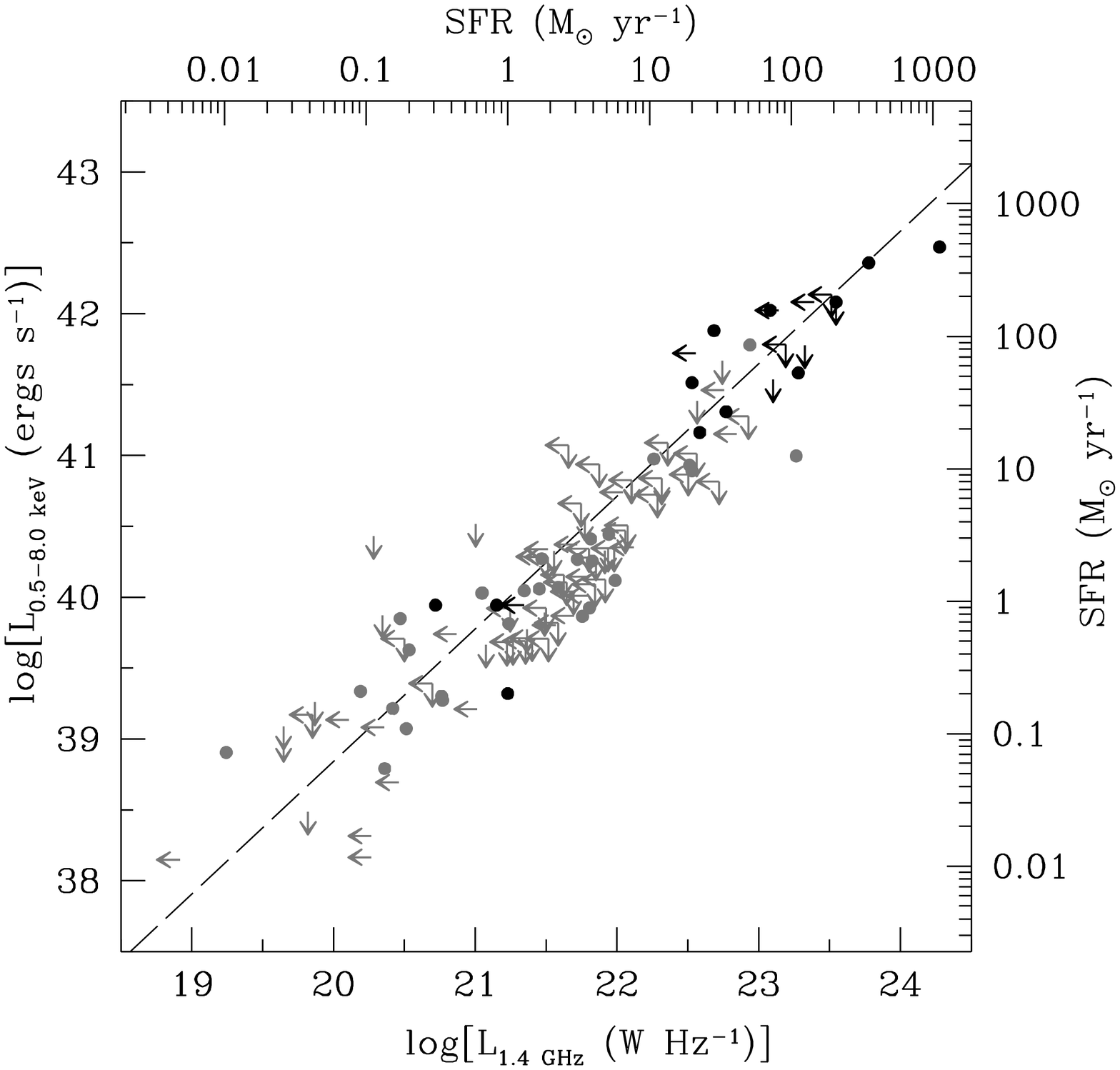}}
\caption{X-ray-radio luminosity comparison for CDF-N (black) and local 
  (grey) sources (see Fig~4 in Bauer et~al. 2002).}}
{\centerline{\includegraphics[width=2in]{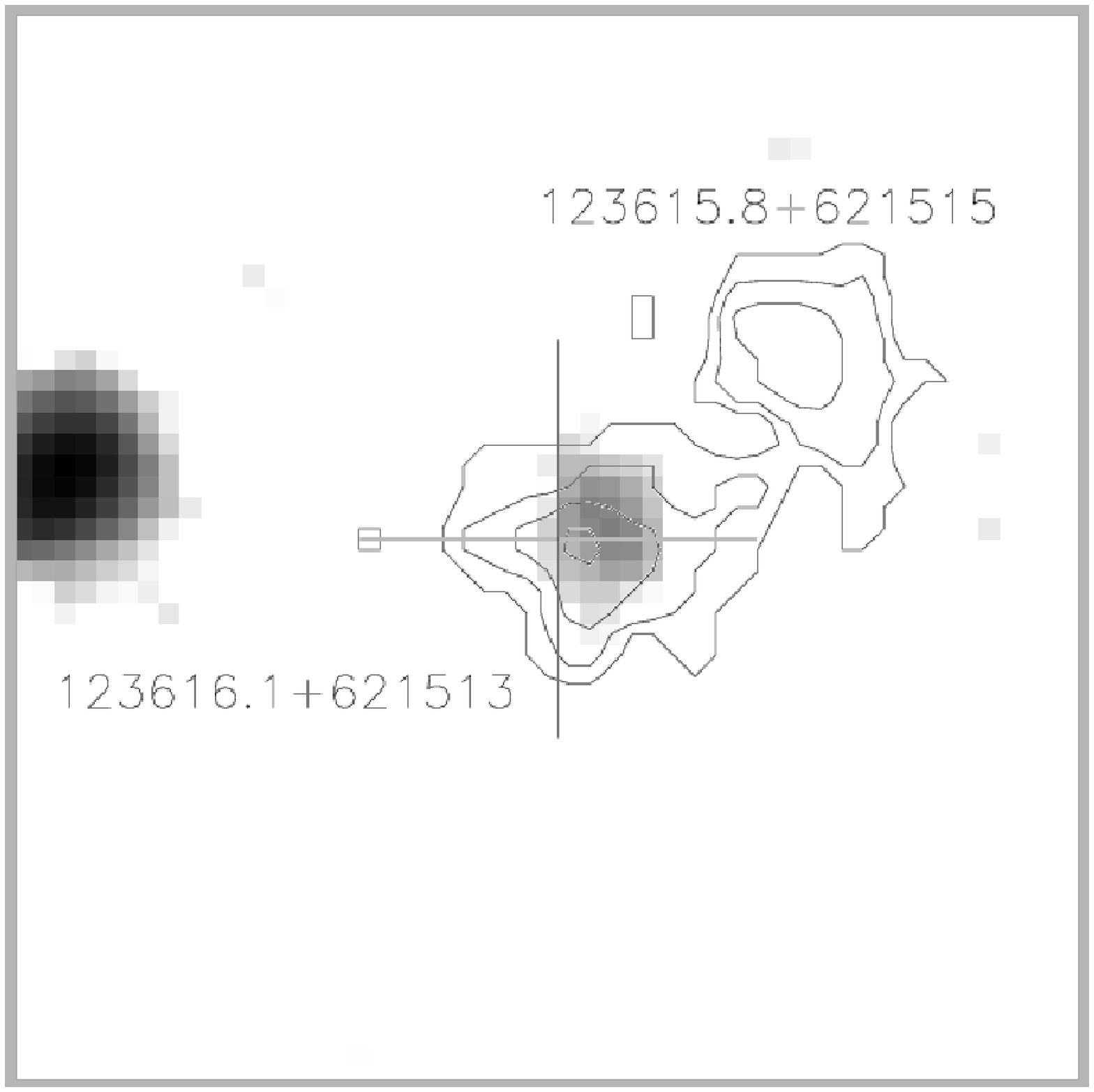}}
\caption{Submm source ($I$-band image; X-ray contours) with likely binary
 AGNs (see Fig~2 in Alexander et~al. 2003b).}}
\end{figure}

%
\section{AGNs in high-redshift submm galaxies}\label{irx}
%

Deep submm surveys have uncovered a population of dust-enshrouded,
luminous galaxies at high-redshift ($z\approx$~1--4; e.g.,\ Smail
et~al.  1997; Hughes et~al. 1998). Both AGN and starburst activity can
theoretically account for the large luminosities of these sources;
however, since few ($<10$\%) submm sources have X-ray counterparts in
moderately deep X-ray surveys, AGNs can only be bolometrically
important if they are Compton thick (e.g.,\ Fabian et~al. 2000;
Hornschemeier et~al. 2000). The CDF-N is sensitive enough to place
direct constraints on the presence and properties of AGNs in submm
galaxies.

Seven bright ($f_{\rm 850\mu m}\ge$~5~mJy; S/N$\ge4$) submm sources
have X-ray counterparts in a 70.3~arcmin$^2$ area centred on the CDF-N
(Alexander et~al. 2003b); using the most recent submm catalog of Borys
et~al. (2003), this corresponds to 54\% of the bright submm sources!
The X-ray emission from five of these sources is clearly AGN
dominated, while the X-ray emission from the other two sources may be
star-formation dominated (Alexander et~al. 2003b). X-ray spectral
analyses of the five AGNs indicate that all are heavily obscured;
however, with 1--2 possible exceptions, the absorption appears to be
Compton thin and the AGNs are of moderate luminosity (Alexander et~al.
2003b). Consequently, the AGNs make a negligible contribution to the
bolometric luminosity. This may imply that the central massive black
holes are in their growth phase.

Interestingly, two ($\approx$~30\%) of the seven submm sources are
individually associated with X-ray pairs (Alexander et~al. 2003b; see
Fig~4). The small angular separations of these pairs
($\approx$~2--3$^{\prime\prime}$) correspond to just $\approx$~20~kpc
at $z=2$ (approximately one galactic diameter); the probability of a
chance association is $<$1\%. We may be witnessing the interaction or
merging of AGNs in these sources (a low-redshift example of this
binary AGN behaviour is NGC~6240; Komossa et~al.  2003). Since only
five ($\approx$~3\%) of the 193 X-ray sources in this region are close
X-ray pairs ($<$3$^{\prime\prime}$ separation), binary AGN behaviour
appears to be closely associated with submm galaxies (see also Smail
et~al. 2003).

%
\begin{acknowledgments}
%
Support came from NSF CAREER award AST-9983783, CXC grant G02-3187A,
Chandra fellowship grant PF2-30021, and the Royal Society. We
thank D. Schneider and C. Vignali for their considerable help in the 
CDF-N project.
\end{acknowledgments}

{

\bibliographystyle{apalike}
\chapbblname{chapbib}
\chapbibliography{logic}

}

\end{document}